\documentclass{PoS}
\usepackage{bbm}

\PoS{PoS(HEP2005)186}

\title{Neutrino mass matrices, texture zeros, and family symmetries}

\ShortTitle{Neutrino mass matrices, texture zeros, and family symmetries}

\author{\speaker{Walter Grimus}\\
        University of Vienna, Austria\\
        E-mail: \email{walter.grimus@univie.ac.at}}

\abstract{
We demonstrate that Abelian family symmetries
allow one to enforce
texture zeros in arbitrary entries
of the fermion mass matrices. Placing
zeros in any number of elements of all occurring mass matrices can be
done with two alternative methods; one of them utilizes the group
$\mathbbm{Z}_n$ with $n$ sufficiently high. Concentrating on the
lepton sector and on neutrino masses, we discuss the
methods in the case of 
seesaw models and scalar triplet models. As an illustration, we present
an example for each type of model.
}

\FullConference{International Europhysics Conference on High Energy Physics\\
		 July 21st - 27th 2005\\
		 Lisboa, Portugal}

\begin{document}

Texture zeros in fermion mass matrices~\cite{fritzsch} present the
simplest procedure to reduce the number of parameters and to induce
relations
among
the physical quantities (masses,
mixing angles,
and CP phases). At first sight this procedure  
is quite arbitrary and in general it will not lead to
renormalizable
models. However,
we have
shown that schemes with
texture zeros can be promoted to
renormalizable
models by an enlargement of the scalar sector~\cite{tex0}:
\emph{For every set of fermion mass matrices with texture zeros in
  arbitrary entries, there exists a scalar sector such that the
  texture zeros are enforced by means of Abelian symmetries.}
In this talk we confine ourselves to the lepton sector with Majorana
neutrinos and extensions of the Standard Model (SM) below the GUT
scale.
However,
we emphasize that our method is completely general
and also applies to the quark sector.

We
will discuss the lepton sector with the seesaw mechanism
and show that there are two methods
for the symmetry implementation of texture zeros.
For this purpose, we
consider the Yukawa Lagrangian~\cite{tex0} 
\begin{equation}\label{LY}
\mathcal{L}_Y = -\sum_{a,b=1}^3 \left( 
\Gamma_{ab} \bar \ell_{Ra} \phi_{ab}^\dagger D_{Lb} +
\tilde\Gamma_{ab}\, \bar \nu_{Ra} {\tilde\phi}_{ab}^\dagger D_{Lb}
+ \frac{1}{2}\, Y_{ab}\, \chi_{ab} \bar \nu_{Ra} C \bar \nu_{Rb}^T
\right) + \mbox{H.c.},
\end{equation}
where $D_L$ denotes the left-handed doublets, $\ell_R$ the
right-handed charged singlets, and $\nu_R$ the right-handed neutrino
singlets. Note that there is one scalar multiplet 
for every fermion bilinear! Thus we have nine 
Higgs doublets $\phi_{ab}$ with hypercharge $+1$, nine Higgs 
doublets ${\tilde\phi}_{ab}$ with hypercharge $-1$, and six gauge
singlet scalars $\chi_{ab} \equiv \chi_{ba}$. 
The corresponding vacuum expectation
values (VEVs) are denoted by $v_{ab}$,
$w_{ab}^\ast$, and $X_{ab}$, respectively.
The charged-lepton mass matrix is given by 
$\left( M_\ell \right)_{ab} = v_{ab}^\ast \Gamma_{ab}$, 
the neutrino Dirac mass matrix by 
$\left( M_D \right)_{ab}= w_{ab} \tilde\Gamma_{ab}$,
and the mass matrix of the heavy neutrino singlets by
$\left( M_R \right)_{ab} = X_{ab} Y_{ab}$.

\paragraph{Method 1:} 
We introduce the Abelian symmetry group 
$\mathcal{G} = \times_f\, \mathcal{G}(f)$ for 
$f = \ell_{Ra},\: D_{La},\: \nu_{Ra}$
($a = 1,2,3$), which has thus 
nine factors. Then, in order to allow the couplings in~(\ref{LY}), the
scalar multiplets transform as 
\begin{equation}
\phi_{ab}: \; \mathcal{G}^*(\ell_{Ra}) \otimes \mathcal{G}(D_{Lb}),
\quad
{\tilde\phi}_{ab}: \; \mathcal{G}^*(\nu_{Ra}) \otimes \mathcal{G}(D_{Lb}),
\quad
\chi_{ab}: \; \mathcal{G}(\nu_{Ra}) \otimes \mathcal{G}(\nu_{Rb}). 
\end{equation}
There is one scalar multiplet for every entry in all three mass
matrices.
Now it is easy to place zeros in arbitrary entries of $M_\ell$,
$M_D$,
and $M_R$. Consider for instance $M_\ell$. If there
exists a 
$\phi_{ab}$ transforming as 
$\mathcal{G}^*(\ell_{Ra}) \otimes \mathcal{G}(D_{Lb})$,
then $\Gamma_{ab} \neq 0$; if such a $\phi_{ab}$ does not
occur, then $\Gamma_{ab} = 0$.

\paragraph{Method 2:} 
We consider the symmetry group
$\mathcal{G} = \mathbbm{Z}_n$ or
$\mathcal{G} = \mathbbm{Z}_n \times \mathbbm{Z}_2$ 
with $n$ sufficiently high. 
It turns out that in the multi-Higgs SM
with the seesaw mechanism
one never needs a larger group~\cite{tex0} than
$\mathbbm{Z}_{12} \times \mathbbm{Z}_2$.
This is easily demonstrated by assuming, 
e.g., that $\bar\ell_R$ and $\bar\nu_R$ transform as 
$(\omega, \omega^2, \omega^5)$ and $D_L$
transform
as $(\omega, \omega^3, \omega^8)$ with $\omega = \exp (i\pi/6)$.
Consequently, fermionic bilinears transform as 
\begin{equation}\label{bilineartrans}
\bar \ell_{Ra} D_{Lb},\,\bar \nu_{Ra} D_{Lb} \sim
\left( \begin{array}{ccc}
\omega^2 & \omega^4 & \omega^9 \\
\omega^3 & \omega^5 & \omega^{10} \\
\omega^6 & \omega^8 & \omega
\end{array} \right), \quad
\bar \nu_{Ra} C \bar \nu_{Rb}^T \sim
\left( \begin{array}{ccc}
\omega^2 & \omega^3 & \omega^6 \\
\omega^3 & \omega^4 & \omega^7 \\
\omega^6 & \omega^7 & \omega^{10}
\end{array} \right).
\end{equation}
The additional 
$\mathbbm{Z}_2:\, {\tilde\phi}_{ab} \to -{\tilde\phi}_{ab}, \:
\nu_R \to -\nu_R$ 
couples the $\phi_{ab}$ solely to $\ell_R$ and the 
${\tilde \phi}_{ab}$ solely to $\nu_R$.
Now the argument for placing zeros goes as before. 
Consider Eq.~(\ref{bilineartrans}); if for instance
$\phi_{13}$ is present transforming 
as $\omega^9$
under $\mathbbm{Z}_{12}$,
then $\left( M_\ell \right)_{13} \neq 0$, and so on.

Some remarks to both methods are at order. 
In practice, in predictive models there are many texture zeros, thus 
rather few scalars are necessary and a proliferation of scalars is avoided.
Moreover, Methods 1 and 2 often merge more or less. The symmetry group 
$\mathcal{G}$ is large, therefore usually 
soft breaking of $\mathcal{G}$ in the scalar potential is necessary 
to avoid Goldstone bosons.

Let us now consider texture zeros in neutrino mass matrices.
We assume a diagonal $M_\ell$, which amounts to six texture zeros in
this matrix. Then, as shown in~\cite{frampton}, for the Majorana mass
matrix $\mathcal{M}_\nu$ of the light neutrinos, there are seven 
viable textures with two zeros. Modulo phase redefinitions, such mass
matrices have five physical parameters. Since there are nine physical
quantities in $\mathcal{M}_\nu$ 
(three neutrino masses, three mixing angles, one CKM-like phase, 
and two Majorana phases), in
such a scenario one has four relations among the physical
quantities~\cite{frampton,xing}. 

As an illustration we consider two cases of~\cite{frampton}:
\begin{equation}\label{cases}
\mbox{Case A$_2$:} \quad \mathcal{M}_\nu \sim 
\left( \begin{array}{ccc} 
0 & \times & 0 \\ \times & \times & \times \\ 0 & \times & \times 
\end{array} \right),
\quad
\mbox{Case C:} \quad
\mathcal{M}_\nu \sim 
\left( \begin{array}{ccc} 
\times & \times & \times \\ \times & 0 & \times \\ \times & \times & 0 
\end{array} \right).
\end{equation}

There are several possible type I seesaw realizations of 
$\mathcal{M}_\nu = -M_D^T M_R^{-1} M_D$ for Case A$_2$,
see~\cite{honda}. As an example we take 
\begin{equation}
M_D \sim \left( \begin{array}{ccc}
\times & 0 & \times \\ 0 & 0 & \times \\ 0 & \times & 0 \end{array} \right),
\quad 
M_R \sim \left( \begin{array}{ccc}
\times & 0 & \times \\ 0 & \times & 0 \\ \times & 0 & 0 \end{array} \right).
\end{equation}
Applying Method 1, we observe that $M_\ell$ being diagonal allows to
make the identification 
$\mathcal{G} \left( \ell_{Ra} \right) \equiv 
\mathcal{G} \left( D_{La} \right)$. We choose 
$\mathcal{G} \left( D_{La} \right) = \mathbbm{Z}_2 \left( D_{La} \right)$,
$\mathcal{G} \left( \nu_{Ra} \right) 
= \mathbbm{Z}_4 \left( \nu_{Ra} \right)$. Then we straightforwardly
arrive at the scalar sector and its transformation properties~\cite{tex0}:
\begin{equation}
\begin{array}{cc}
\tilde\phi_{11}: \, \mathbbm{Z}^*_4 \left( \nu_{R1} \right)
\otimes \mathbbm{Z}_2 \left( D_{L1} \right), & \hphantom{xx}
\chi_{11}: \, \mathbbm{Z}_4 \left( \nu_{R1} \right) 
\otimes \mathbbm{Z}_4 \left( \nu_{R1} \right), \\
\tilde\phi_{13}: \, \mathbbm{Z}^*_4 \left( \nu_{R1} \right)
\otimes \mathbbm{Z}_2 \left( D_{L3} \right), & \hphantom{xx}
\chi_{22}: \, \mathbbm{Z}_4 \left( \nu_{R2} \right) 
\otimes \mathbbm{Z}_4 \left( \nu_{R2} \right), \\
\tilde\phi_{23}: \, \mathbbm{Z}^*_4 \left( \nu_{R2} \right)
\otimes \mathbbm{Z}_2 \left( D_{L3} \right), & \hphantom{xx}
\chi_{13}: \, \mathbbm{Z}_4 \left( \nu_{R1} \right)
\otimes \mathbbm{Z}_4 \left( \nu_{R3} \right), \\
\tilde\phi_{32}: \, \mathbbm{Z}^*_4 \left( \nu_{R3} \right)
\otimes \mathbbm{Z}_2 \left( D_{L2} \right). &
\end{array}
\end{equation}
In addition, in the charged-lepton sector one Higgs doublet
transforming trivially is needed. Thus we end up with five Higgs
doublets and three scalar singlets.
In this example we have seen 
the typical simplification of Method 1 if one 
mass matrix happens to be diagonal: $\mathcal{G}$ reduces to a
direct product of only six groups. 
Applying Method 2, we find a more economical symmetry realization of
Case A$_2$: $\mathcal{G} = \mathbbm{Z}_8$ with two Higgs doublets and
two scalar singlets~\cite{tex0}.

All cases found in~\cite{frampton} can also be realized via Abelian
symmetries and scalar triplets~\cite{GL-2tex0}; no right-handed
neutrino singlets are needed and there is a single Higgs doublet $\phi$ 
responsible for the charged lepton masses. 
We exemplify this with Case~C.\footnote{In~\cite{frigerio}, this case is
  realized via the non-Abelian symmetry group $\mathbbm{Q}_8$.} 
The Yukawa couplings of the scalar triplets are given by 
\begin{equation}\label{LYDelta}
\mathcal{L}_{Y\,\Delta} =
\frac{1}{2}\, \sum_j \sum_{a,b=1}^3
h_{ab}^j\, D_{La}^\mathrm{T} C^{-1}
\left( i \tau_2 \Delta_j \right) D_{Lb}
+ \mbox{H.c.}
\end{equation}
The VEVs $w_j$ of the neutral components of the scalar triplets
$\Delta_j$ generate the neutrino mass matrix 
$\mathcal{M}_\nu = \sum_j w_j\, h^j$.
We confine ourselves to Method 2 and make the ansatz
\begin{equation}
D_{La} \to p_a D_{La},
\quad
\ell_{Ra} \to p_a \ell_{Ra},
\quad
\phi \to \phi
\quad \mbox{with} \quad 
\left| p_a \right| = 1
\end{equation}
for the symmetry transformation.
With all phase factors $p_a$ different from each other, $M_\ell$ is
automatically diagonal. 
A suitable choice is
$p_e = 1,\ p_\mu = i,\ p_\tau = -i$.
Then the transformation properties of the bilinears in leptonic
doublets determine the number and the transformation properties of the
scalar triplets: 
\begin{equation}
D_{La}^T C^{-1} D_{Lb} \sim
\left( \begin{array}{ccc}
1 & i & -i \\ i & -1 & 1 \\ -i & 1 & -1
\end{array} \right)
\Rightarrow \left\{
\begin{array}{ccc} 
\Delta_1 & \to & \Delta_1, \\
\Delta_2 & \to & -i \Delta_2,  \\
\Delta_3 & \to & i \Delta_3.
\end{array}
\right.
\end{equation}
Thus we have found a symmetry realization of Case~C with 
the family symmetry $\mathbbm{Z}_4$ which needs 
only one Higgs doublet and three scalar triplets.
For the symmetry realization of all cases of~\cite{frampton}
with scalar triplet models see~\cite{GL-2tex0}.
Texture zeros in $\mathcal{M}_\nu$ with triplet realizations are
stable under the renormalization group running because only one Higgs
doublet is present.

In summary, the two methods presented in~\cite{tex0} allow to embed
all kinds of fermion mass matrix schemes with texture zeros in
renormalizable models,
possibly at the cost of a proliferation of the scalar sector;
such an embedding is not unique.
We emphasize the versatility of the methods which would equally well
apply in the quark sector or in Grand Unified Theories.
Finally we note that not only texture zeros in $\mathcal{M}_\nu$ but
also in $\mathcal{M}_\nu^{-1}$ can lead to interesting
models~\cite{inverse}.


\begin{thebibliography}{9}

\bibitem{fritzsch}
H.~Fritzsch, Z.-Z.~Xing, 
\emph{Mass and flavor mixing schemes of quarks and leptons},
\emph{Prog. Part. Nucl. Phys.} {\bf 45} (2000) 1 
[{\tt hep-ph/9912358}].

\bibitem{tex0}
W.~Grimus, A.S.~Joshipura, L.~Lavoura, M.~Tanimoto,
\emph{Symmetry realization of texture zeros},
\emph{Eur. Phys. J.} {\bf C 36} (2004) 227
[{\tt hep-ph/0405016}].

\bibitem{frampton}
P.H.~Frampton, S.L.~Glashow, D.~Marfatia, 
\emph{Zeroes of the neutrino mass matrix},
\emph{Phys. Lett.} {\bf B 536} (2002) 79 
[{\tt hep-ph/0201008}].

\bibitem{xing}
Z.-Z.~Xing, 
\emph{Texture zeros and Majorana phases of the neutrino mass matrix},
\emph{Phys. Lett.} {\bf B 530} (2002) 159 
[{\tt hep-ph/0201151}]; \\
Z.-Z.~Xing, 
\emph{A full determination of the neutrino mass spectrum from two-zero
textures of the neutrino mass matrix},
\emph{Phys. Lett.} {\bf B 539} (2002) 85 
[{\tt hep-ph/0205032}].

\bibitem{honda}
M.~Honda, S.~Kaneko, M.~Tanimoto, 
\emph{Seesaw enhancement of bi-large mixing in two-zero textures},
\emph{Phys. Lett.} {\bf B 593} (2004) 165 
[{\tt hep-ph/0401059}].

\bibitem{GL-2tex0}
W.~Grimus, L.~Lavoura,
\emph{On a model with two zeros in the neutrino mass matrix},
\emph{J. Phys.} {\bf G 31} (2005) 693 
[{\tt hep-ph/0412283}].

\bibitem{frigerio}
M.~Frigerio, S.~Kaneko, E.~Ma, M.~Tanimoto,
\emph{Quaternion family symmetry of quarks and leptons},
\emph{Phys. Rev.} {\bf D 71} (2005) 011901 
[{\tt hep-ph/0409187}].

\bibitem{inverse}
L.~Lavoura,
\emph{Zeros of the inverted neutrino mass matrix},
\emph{Phys. Lett.} {\bf B 609} (2005) 317 
[{\tt hep-ph/0411232}]; \\
W.~Grimus, S.~Kaneko, L.~Lavoura, H.~Sawanaka, M.~Tanimoto,
\emph{$\mu$--$\tau$ antisymmetry and neutrino mass matrices},
hep-ph/0510326.

\end{thebibliography}
\end{document}